\documentclass[letterpaper,journal]{IEEEtran}
\usepackage{amsmath,amsfonts}
\usepackage{algorithmic,algorithm,array}
\usepackage[caption=false,font=normalsize,labelfont=sf,textfont=sf]{subfig}
\usepackage{textcomp,stfloats,graphicx}
\usepackage{url}
\usepackage{verbatim}
\usepackage{cite}
\usepackage{multirow}
\usepackage{booktabs}
\usepackage{fancyhdr}

\begin{document}
\title{Token Communication-Assisted Collaborative Embodied Artificial Intelligence: Concepts, Framework, and Opportunities}
\author{\IEEEauthorblockN{Peng Yi, and Ying-Chang Liang,~\IEEEmembership{Fellow, IEEE}} \\ 
\thanks{Peng Yi is with the National Key Laboratory of Wireless Communications, and also with the Center for Intelligent Networking and Communications (CINC), University of Electronic Science and Technology of China (UESTC), Chengdu 611731, China (email: yipengcd@outlook.com).
\par Ying-Chang Liang is with the Institute of Fundamental and Frontier Sciences and the Center for Intelligent Networking and Communications (CINC), University of Electronic Science and Technology of China (UESTC), Chengdu 611731, China (email: liangyc@ieee.org). The corresponding author is Y.-C. Liang.}
}

\maketitle
\thispagestyle{fancy}
\fancyhf{}
\renewcommand{\headrulewidth}{0pt}
\lhead{\parbox{\textwidth}{\scriptsize This work has been submitted to the IEEE for possible publication. Copyright may be transferred without notice, after which this version may no longer be accessible.}}

\begin{abstract}
Collaborative embodied artificial intelligence (CEAI) enables multiple physical agents to perceive, reason, and act cooperatively in dynamic environments. Effective communication is essential for CEAI, yet CEAI agents must exchange not only large multimodal observations but also task-relevant insights, intents, and interactive information over long horizons. This article investigates token communication (TokCom) as a native intelligence interface for CEAI, in which tokens serve jointly as compact semantic carriers for communication and fundamental inference units for generative foundation models (GFMs). We first discuss how TokCom supports insight sharing, intent alignment, and interactive control among embodied agents. We then propose a TokCom-assisted CEAI framework driven by a task-adaptive communication protocol. Comprising a compact codebook, syntax rules, and contextual examples, this protocol guides GFM-based transceivers to distill messages into compact tokens and reconstruct them after wireless transmission. A case study on collaborative object transport demonstrates that the proposed TokCom framework substantially reduces the source payload bit consumption while preserving task efficiency and showing robustness under noisy channels. Finally, we outline future research directions.
\end{abstract}


\section{Introduction}
Collaborative embodied artificial intelligence (CEAI) represents a decentralized paradigm in which multiple physical agents operate within a shared environment, coordinating their actions to achieve common goals~\cite{IJCAI/25/Wu}. 
Due to partial observability and inherent physical constraints of individual agents, communication is essential for them to exchange perceptions and intents, enabling coordinated actions to effectively accomplish complex tasks~\cite{CVPR/19/Jain,COMST/26/ZhangP}. 
To fulfill the vision of CEAI, next-generation wireless systems must evolve beyond supporting conventional human-to-human (H2H) communication to enable intelligent multi-agent communication, reasoning, and coordination~\cite{COMST/26/ZhangR}.

However, existing communication paradigms are insufficient to support the demands of CEAI. In particular, conventional communication systems are designed to ensure reliable bit-level transmissions, which become inefficient for embodied agents, as faithfully transmitting their continuously collected multimodal raw data would consume excessive bandwidth and incur high latency~\cite{CVPR/Liu}. Semantic communication (SemCom) alleviates part of this burden by extracting and transmitting task-relevant meaning, thereby reducing redundant source information before transmission~\cite{COMST/24/Lu}. Nevertheless, existing SemCom systems are developed for predefined tasks. This makes it difficult to apply them directly to dynamic CEAI scenarios, in which embodied agents must continuously transform multimodal observations into compact messages and exchange them to support general, adaptive, and long-horizon multi-agent collaboration. In such settings, communication is not limited to conveying extracted data semantics, but also involves sharing intentions, task states, environmental changes, requests, and coordination constraints over long interaction horizons.

\begin{figure*}
  \centering
  \includegraphics[width=0.9\textwidth]{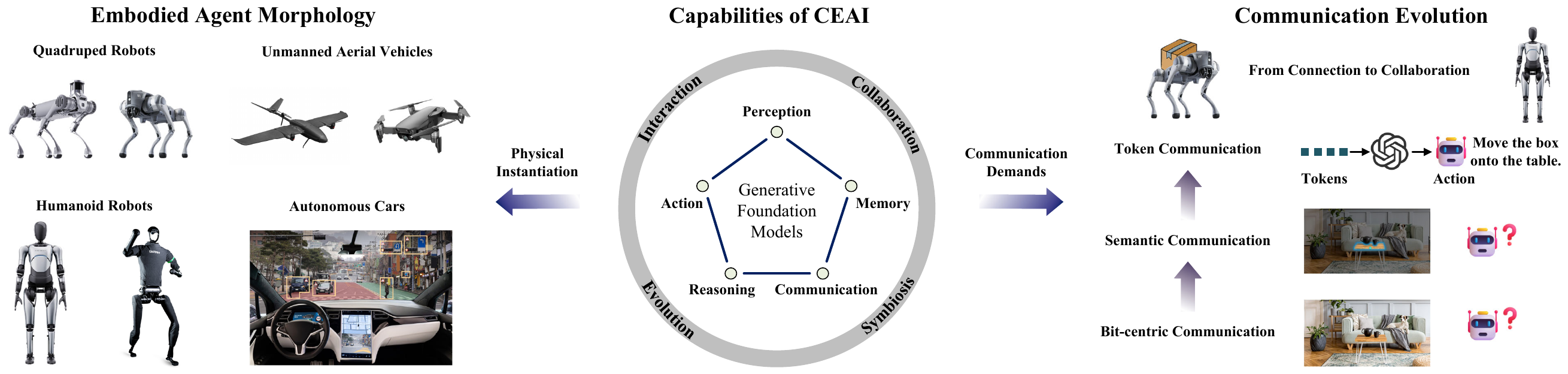}
  \caption{A conceptual framework for CEAI. The figure illustrates three key components. Left: A diverse array of embodied intelligent agents. Center: The core capabilities of embodied intelligence, anchored by GFMs. These capabilities, including perception, memory, reasoning, communication, and action, drive advanced behaviors such as environmental interaction, multi-agent collaboration, and continuous evolution, ultimately fostering symbiosis with the physical world. Right: The requisite evolution of communication paradigms, shifting the focus from simple connectivity to intelligent collaboration.}
  \label{fig:CEAI}
\end{figure*}

Driven by the rapid evolution of generative foundation models (GFMs) for artificial intelligence (AI)-native wireless systems, token communication (TokCom) has become a potential approach to enable CEAI~\cite{MWC/25/Qiao}. In this paradigm, tokens serve concurrently as compact carriers of multimodal semantics and as the fundamental inference units of GFMs. This dual role is particularly appealing for CEAI.
On the one hand, as compact semantic carriers, tokens encode heterogeneous data from various modalities into a unified format, providing a common representation that streamlines information transmission among embodied agents. 
On the other hand, as the fundamental inference units of GFMs, they contain essential semantics that can be directly processed for reasoning, so that low-level exchanges can support intention alignment, plan negotiation, and collaborative decision-making.
In this sense, TokCom goes beyond simple data representation and serves as a native intelligence interface for multi-agent collaboration at the intelligent edge.

Despite these advantages, applying TokCom to CEAI over wireless networks presents significant challenges. 
First, current GFMs are inherently designed for open-ended reasoning and generation, whereas TokCom among embodied agents must be significantly more concise for specific tasks. In GFMs, tokenizers typically rely on an expansive general vocabulary, often containing on the order of $10^5$ entries. With a vocabulary size of $V$, transmitting a single token index necessitates $\lceil \log_2 V \rceil$ source bits, corresponding to $17$ or $18$ bits per token when $V \approx 10^5$. However, for a specific embodied task, only a highly restricted subset of semantic concepts, actions, intentions, and coordination commands is typically required. Consequently, the massive generic vocabulary can be compressed into a streamlined, task-adaptive agent language, enabling agents to exchange information with significantly higher efficiency over wireless channels. 
Second, unlike traditional wireless systems that focus on point-to-point metrics such as bit error rate and single-shot task accuracy, CEAI prioritizes long-horizon collaborative task completion. Therefore, the performance of wireless TokCom should be evaluated by how transmitted tokens affect reasoning, coordination, and task success, rather than by token recovery accuracy alone. This calls for a paradigm shift toward effectiveness-level metrics that directly connect wireless transmission quality with the final outcomes of embodied multi-agent collaboration.

Motivated by these challenges, this article investigates TokCom-assisted CEAI. The main contributions are summarized as follows: 
\begin{itemize}
  \item We provide an overview of CEAI and its associated communication paradigms, analyze the limitations of conventional bit-centric and SemCom paradigms, and introduce the emerging TokCom paradigm. Furthermore, we highlight how TokCom assists CEAI from three distinct perspectives: insight sharing, intent alignment, and interactive control.
  \item We propose a TokCom-assisted CEAI framework in which GFMs initialize a task-adaptive token protocol to align communication vocabularies among collaborating agents. During transmission, GFM-based encoders distill natural-language messages into compact token sequences for binary transmission before channel coding, and GFM-based decoders reconstruct them after reception, thereby facilitating efficient collaboration over wireless channels.
  \item We evaluate the proposed framework through a multi-agent collaborative object-transport case study. By assessing both long-horizon task effectiveness and communication efficiency, we show that the proposed framework substantially reduces source payload bit consumption while maintaining robust collaborative performance.
\end{itemize}

\section{Overview of CEAI and TokCom}
This section first presents CEAI from the perspectives of representative embodied agents, application scenarios, and core capabilities, as illustrated in Fig.\,\ref{fig:CEAI}. It then revisits the role of communications in CEAI, analyzes the limitations of existing bit-centric and SemCom paradigms, and finally introduces TokCom as a promising communication paradigm. 

\subsection{CEAI}
CEAI is a decentralized paradigm in which multiple physical agents operate in a common environment, coordinating their actions to achieve collective goals. Different from passive AI, which relies on predictions over independent observations of the world, CEAI comprises active, embodied agents that deliberately perceive and interact with their environments to pursue explicit goals~\cite{MCAS/25/Feng}. The defining distinction is embodiment, in which CEAI agents are physically instantiated, taking diverse forms depending on their target applications. Representative morphologies include:
\begin{itemize}
  \item \textbf{Quadruped Robots}: These four-legged robots are designed for ground locomotion, capable of traversing complex and varied terrains. 
  \item \textbf{Humanoid Robots}: With a human-like form and inherent adaptability, these versatile robots can be deployed in a wide range of applications, often to perform tasks traditionally done by humans. 
  \item \textbf{Unmanned Aerial Vehicles (UAVs)}: Categorized as either fixed-wing or rotary-wing, UAVs perceive their surroundings using cameras, light detection and ranging (LiDAR), and other sensors, adjusting their flight paths via propeller control. 
  \item \textbf{Autonomous Vehicles}: These agents exemplify the core principles of embodied intelligence by perceiving the dynamic traffic environment, making real-time decisions, and physically controlling the vehicle. 
\end{itemize}

These embodied agents support a wide range of applications. In smart manufacturing, they serve as mobile inspectors and precision operators for production monitoring, collaborative assembly, and hazardous material handling. In intelligent transportation, UAVs support aerial logistics, while autonomous vehicles enable passenger transport and safer road mobility. For public services, robots for service, inspection, and rescue assist in healthcare, infrastructure monitoring, environmental protection, and disaster response.

Despite these diverse physical forms and applications, the core of CEAI is anchored by emerging GFMs~\cite{MCAS/25/Feng}. Pretrained on large-scale multimodal data, GFMs provide the reasoning and generative capabilities that support the following core functions of CEAI agents: 
\begin{itemize}
  \item \textbf{Perception}: Agents perceive their environment through a range of modalities, such as visual, auditory, and tactile signals, and then interpret these observations to support memory, reasoning, communication and action. 
  \item \textbf{Memory}: Agents maintain and access a repository of information that includes, but is not limited to, their internal states, environmental contexts, and experiences.
  \item \textbf{Reasoning}: Leveraging perception and memory, agents draw inferences, formulate plans, make decisions, and solve complex problems. 
  \item \textbf{Communication}: Agents exchange information and coordinate with other agents to achieve collective goals. 
  \item \textbf{Action}: Agents execute physical tasks within their environment, such as navigation, grasping, and object manipulation. 
\end{itemize} 

By combining these capabilities into a closed loop, CEAI translates physical interactions and multi-agent collaborations into experience-driven memory updates, continually evolving the agents for progressively deeper engagement with the physical world.

\subsection{Communications in CEAI} 
Although individual agents possess diverse capabilities, they remain limited by partial observability and inherent physical boundaries. To overcome these individual limitations and tackle complex tasks requiring collective effort, effective multi-agent communication is indispensable~\cite{MCOM/26/Chen}. It serves as the key mechanism that enables agents to \textit{share perceptions}, \textit{exchange intentions}, and \textit{coordinate actions}, transforming isolated entities into a collaborative system. 
Generally, multi-agent communication research has evolved along two distinct paradigms: 
\begin{itemize} 
  \item \textbf{Bit-Centric Communication Paradigm:} Rooted in the classical Shannon framework, this approach prioritizes reliable bit transmissions. However, for embodied agents equipped with data-intensive sensors, this strategy is often inefficient, necessitating the transmission of massive raw data volumes even for simple tasks~\cite{CVPR/Liu}. Furthermore, the ``cliff effect,'' characterized by a sharp performance degradation under poor channel conditions, can be catastrophic for collaboration, particularly in complex electromagnetic environments. As a result, existing studies under this paradigm typically formulate the problem in terms of channel capacity or point-to-point reliability, without explicitly capturing how transmission errors propagate to subsequent collaborative processes.
  \item \textbf{SemCom Paradigm:} Shifting the focus from data fidelity to semantic fidelity, this paradigm prioritizes the accurate conveyance of meaning. By extracting and transmitting only essential, high-level information, it enhances spectral efficiency and resilience against channel noise~\cite{TCCN/26/Jiang}. Despite these advantages, current SemCom systems are typically highly specialized for specific tasks. Because their neural weights are statically bound to predefined downstream objectives, these systems require prohibitive retraining whenever the collaborative task changes. Consequently, they lack the generality required for CEAI, in which agents must process heterogeneous data modalities and execute diverse, complex collaborative tasks. 
\end{itemize}

While these paradigms provide a foundation for current communication systems, they remain insufficient for CEAI, as communications in embodied multi-agent systems must address the following challenges: 
\begin{itemize} 
  \item \textbf{Multimodal Grounding:} Embodied agents perceive the physical world through multimodal data from heterogeneous sensors. Consequently, communications must represent and transmit information grounded in multiple modalities. Crucially, this requires cross-modal alignment capabilities, such as translating visual observations into linguistic descriptions or converting language instructions into executable actions. 
  \item \textbf{Task-Oriented Effectiveness:} CEAI prioritizes long-horizon task completion over point-to-point transmission performance. From the Shannon-Weaver perspective, this requirement corresponds to the effectiveness level of communication, namely whether the received information leads to desired actions and outcomes. Accordingly, at the effectiveness level, the value of transmitted information should be measured by its ultimate contribution to collaborative task success, rather than solely by semantic fidelity~\cite{TVT/26/Su}.
\end{itemize}

\subsection{TokCom}
In Transformer-based GFMs, tokens are the elementary units for sequence modeling and are internally mapped to embedding vectors. Text tokens are typically discrete vocabulary indices, whereas visual, audio, or control-related tokens may be represented either as discrete codebook entries or as continuous embeddings. Through tokenization, raw multimodal data are converted into token sequences that preserve task-relevant features and semantics, making them suitable for processing by GFMs.

TokCom represents a paradigm shift by leveraging these tokens as the primary information carriers~\cite{MWC/25/Qiao}. Unlike traditional paradigms that transmit raw signals or engineered features, agents in this framework leverage GFMs to generate tokens for transmission and to directly process received tokens for downstream reasoning and action. TokCom offers distinct advantages for CEAI: 
\begin{itemize}
  \item \textbf{Unified Multimodal Representation:} With appropriate tokenizers, data from diverse modalities can be projected into a unified representation space~\cite{WCL/26/Wei}. This equips TokCom with a natural ability to support the transmission, fusion, and reasoning of multimodal information within a common framework. 
  \item \textbf{Direct Task-Level Utility:} Received tokens can be directly used by the receiving agent's GFM to infer intentions and generate corresponding actions. This closes the loop from communications to execution, allowing communication effectiveness to be evaluated directly by collaborative task success. 
\end{itemize}

\begin{figure*}
  \centering
  \includegraphics[width=0.9\textwidth]{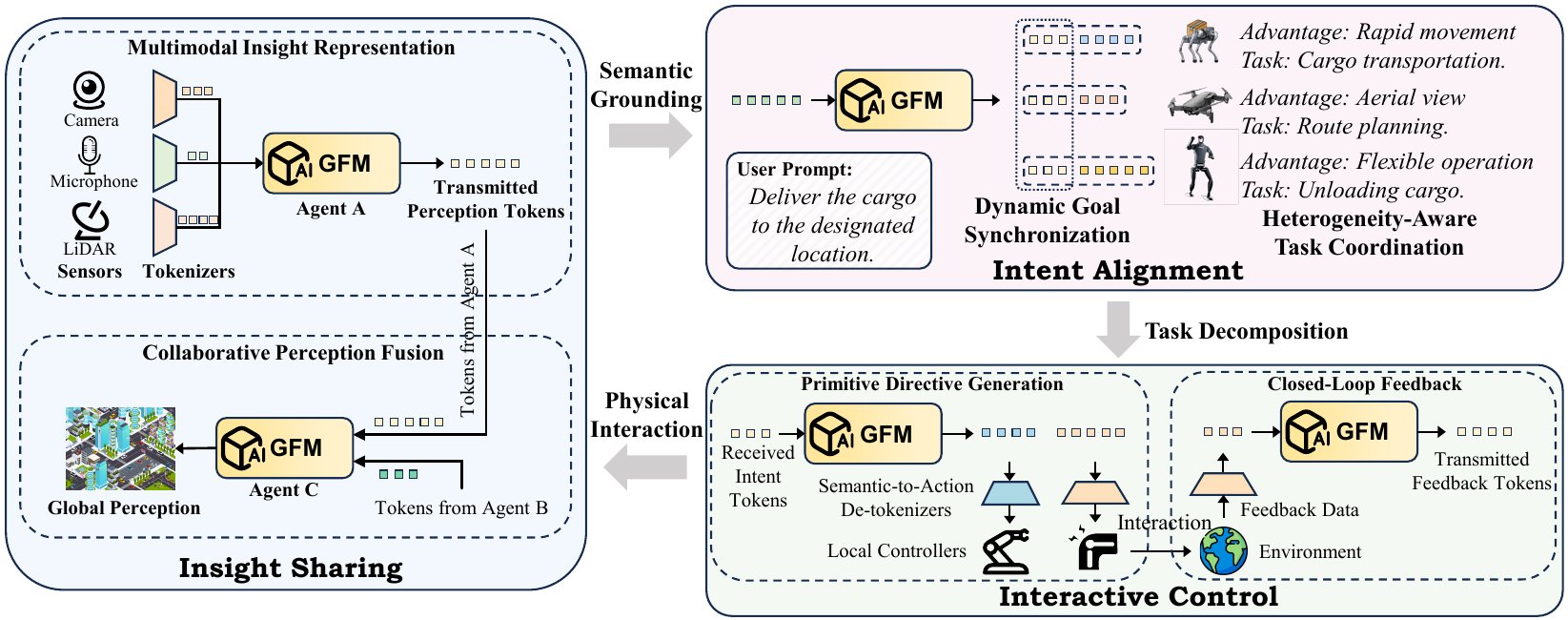}
  \caption{TokCom capabilities in CEAI across insight sharing, intent alignment, and interactive control.}
  \label{fig:assistance}
\end{figure*}

\subsection{Summary}
Based on the overview of CEAI and TokCom, we highlight the following critical insights: 
\begin{itemize}
  \item \textbf{CEAI necessitates effective communications.} Embodied agents are physically distributed and hindered by partial observability and local physical constraints. Thus, effective communications are indispensable for sharing perceptions, exchanging intentions, and coordinating actions. 
  \item \textbf{Existing communication paradigms fall short.} Hindered by bit-centric inefficiencies and the task-specific limitations of current SemCom, existing paradigms struggle in the highly dynamic environments of CEAI. Consequently, neither approach fully supports the multimodal-grounded and task-oriented communication essential for embodied multi-agent collaboration.
  \item \textbf{TokCom emerges as a promising agent language for CEAI.} 
  By encapsulating multimodal observations, high-level semantics, and intentions into tokens that GFMs directly generate and process, TokCom provides a unified interface for embodied collaboration. More importantly, it directly aligns communication effectiveness with collaborative task success.
\end{itemize}

\section{TokCom Assistance in CEAI}
\label{sec:assistance}

In this section, we discuss how TokCom supports CEAI collaboration across three layers: insight sharing at the perception layer, intent alignment at the reasoning layer, and interactive control at the execution layer, as illustrated in Fig.\,\ref{fig:assistance}.

\subsection{Insight Sharing}

TokCom first helps agents overcome sensory heterogeneity and partial observability by transforming local observations into shareable task-relevant insights.

\begin{itemize}
    \item \textbf{Multimodal Insight Representation:}
    Embodied agents rely on heterogeneous sensors and may observe different aspects of the same environment. By representing sensory information as tokens, TokCom provides a common interface through which agents can exchange observations across modalities, such as visual, auditory, LiDAR, or tactile information. This enables agents to build a shared semantic ground for collaboration.

    \item \textbf{Collaborative Perception Fusion:}
    TokCom also supports the fusion of distributed perception. A transmitting agent can distill local observations into receiver-relevant token sequences, reducing the need to transmit raw sensory data. The receiving agent can then integrate tokens from multiple peers with its own local memory, forming a more complete understanding of the environment under bandwidth-limited conditions.
\end{itemize}

\subsection{Intent Alignment}

Beyond sharing insights, effective CEAI requires agents to align their high-level intents prior to interactive control. Distinct from low-level control commands, these abstract intents encompass global team objectives and local task plans. TokCom encapsulates such intents into concise token sequences, facilitating alignment at two complementary levels: synchronizing global goals and negotiating local roles.

\begin{itemize}
    \item \textbf{Dynamic Goal Synchronization:}
    At the team level, task contexts and global objectives may evolve in dynamic physical environments. TokCom allows a cloud server or an elected coordinator to encode updated global goals into token sequences. This synchronizes the collective intent across distributed agents, ensuring that their local reasoning remains aligned with the latest team-level mission.

    \item \textbf{Heterogeneity-Aware Task Coordination:}
    At the individual level, CEAI agents have heterogeneous capabilities in mobility, sensing range, battery level, payload capacity, and manipulation. By exchanging tokens that express their physical constraints and planned subtasks, agents can negotiate role allocations. This aligns their local intents to synthesize complementary plans, thereby preventing conflicting decisions and assigning subtasks to the most suitable agents.
\end{itemize}

\begin{figure*}
  \centering
  \includegraphics[width=0.95\textwidth]{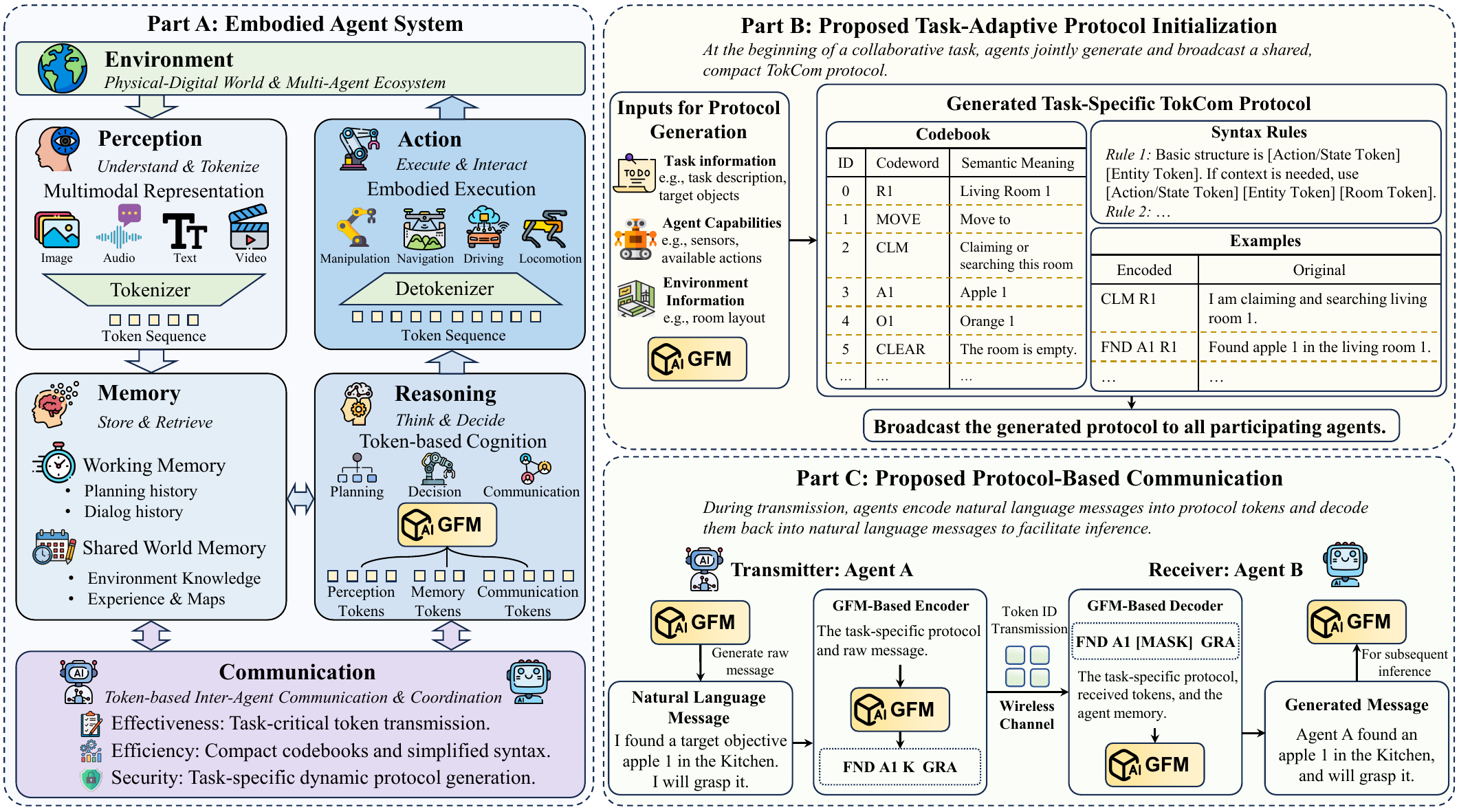}
  \caption{Proposed TokCom-assisted CEAI framework. Part A presents the embodied agent system, including perception, memory, reasoning, action, and communication modules. The proposed components are highlighted in Parts B and C: Part B illustrates task-adaptive protocol initialization, while Part C presents protocol-based communication, in which agents encode natural-language messages into protocol tokens and decode the received tokens for subsequent inference.}
  \label{fig:framework}
\end{figure*}

\subsection{Interactive Control}

At the execution layer, TokCom supports interaction-oriented control by exchanging compact directives and feedback, while leaving low-level actuation to local controllers.

\begin{itemize}
    \item \textbf{Primitive Directive Generation:}
    TokCom can transmit compact tokens that represent parameterized action primitives, such as approach, grasp, or move. Rather than streaming raw control commands, the transmitting agent communicates an abstract directive, and the receiver instantiates it using its local planner, memory, and embodiment-specific controller. This reduces communication overhead while preserving interoperability among heterogeneous agents.

    \item \textbf{Closed-Loop Feedback:}
    Physical cooperation also requires timely feedback about execution states. Agents can exchange lightweight tokens summarizing tracking deviation, contact status, local feasibility, or assistance requests. Such feedback helps partners adjust primitives, change strategies, or coordinate physical interaction without continuously transmitting raw sensor or control signals.
\end{itemize}

\subsection{Summary}
The discussion above highlights how tokens act as a universal semantic carrier across the perception, reasoning, and execution layers of CEAI. By providing a unified representation for insights, intents, and interactions, TokCom overcomes the heterogeneity of agent hardware and modalities, enabling a fully closed-loop collaborative system. However, realizing this potential in practical deployments requires overcoming the inherent mismatch between GFM tokenization and physical wireless channels, a challenge we address in the following section.

\section{Proposed TokCom-Assisted CEAI Framework}
As shown in Part A of Fig.\,\ref{fig:framework}, each CEAI agent consists of five interconnected modules, namely perception, memory, reasoning, communication, and action, all built around a GFM core. Among them, the communication module is central to collaboration, as it enables agents to share observations, align intentions, and coordinate actions. In wireless scenarios, however, communication must preserve task-relevant semantics while remaining efficient under bandwidth limitations and channel fading. This motivates the TokCom-assisted CEAI framework proposed in this section.

\subsection{Motivation}
While Section~\ref{sec:assistance} showed that TokCom can support insight sharing, intent alignment, and interactive control in CEAI, directly transmitting token sequences generated by GFMs poses severe challenges for practical wireless deployment. In real-world scenarios, distributed CEAI agents must communicate over wireless links characterized by stringent bandwidth constraints and time-varying channel fading. However, pretrained GFM tokenizers are fundamentally optimized for inference and generation rather than communication. Being agnostic to wireless constraints, they lack mechanisms to minimize transmission payloads or ensure semantic robustness against channel impairments. This mismatch between inference-oriented tokenization and wireless transmission requirements leads to two major inefficiencies:

\begin{itemize}
  \item \textbf{Sequence-Level Redundancy and Ambiguity:}
  At the sequence level, GFM-generated natural language tends to retain redundant lexical, syntactic, or multimodal details that are unnecessary for physical execution, thereby inflating the transmission payload~\cite{TCCN/26/Jiang}. Furthermore, the same physical phenomenon or task intent can be expressed by highly varied token sequences. This descriptive inconsistency risks introducing ambiguity, as receiving agents may interpret the varied sequences differently, hindering precise coordination.

  \item \textbf{Vocabulary-Level Overhead and Inefficiency:} 
  At the vocabulary level, existing TokCom systems typically transmit token indices directly from a pretrained GFM tokenizer~\cite{MWC/25/Qiao}. Such tokenizers rely on general-purpose, large-scale vocabularies. For example, Gemma 4 employs a vocabulary of approximately $262$k entries, requiring $\lceil\log_2(262\text{k})\rceil=18$ bits to encode a single token index. While acceptable for local inference, this vocabulary-level overhead is prohibitive for frequent multi-agent communication under stringent bandwidth constraints.
\end{itemize}

These observations suggest that pretrained GFM tokenizers are effective for inference and generation, but not well suited to wireless communication and collaboration. CEAI instead requires a task-adaptive and compact agent language that preserves essential semantics while reducing ambiguity and transmission cost.

\subsection{Proposed Framework}
To address the above challenges, we propose a TokCom-assisted CEAI framework, illustrated in Parts B and C of Fig.\,\ref{fig:framework}. The central idea is to insert a task-adaptive protocol layer between GFM reasoning and wireless transmission, translating reasoning-oriented natural language into a compact, transmission-efficient agent language. 
Specifically, the agents first establish a shared task-adaptive communication protocol, and then use this protocol to guide semantic compression and reconstruction during message exchange. In this way, communication is shifted from general-purpose to task-oriented token transmission, thereby improving both efficiency and robustness over wireless links.

\subsubsection{Task-Adaptive Protocol Initialization}
\label{subsec:protocol}
Before a collaborative task begins, a cloud server or an elected coordinator can use a GFM to generate a shared, task-adaptive TokCom protocol and distribute it to all agents. Recent GFM-driven agent systems have begun to adopt explicit communication protocols, such as the model context protocol (MCP) and the agent-to-agent (A2A) protocol, to structure interactions among agents, tools, and environments~\cite{ARXIV/25/Kong}. In contrast to such general software-level protocols, the proposed TokCom protocol is tailored to wireless CEAI by defining task-adaptive semantic primitives, syntax rules, and compact protocol-token IDs. 
Rather than relying on a universal fixed protocol, this protocol can be generated or adapted according to the task information, agent capabilities, and environmental context. It therefore serves as a task-oriented communication interface rather than a general-purpose one. In the following, we refer to the semantic primitives defined by this protocol as \textit{protocol tokens}. Each protocol token is assigned a compact protocol-token identifier (ID) for binary transmission, distinguishing it from the general-purpose GFM tokens produced by pretrained tokenizers. These protocol tokens are generated for the current task and environment rather than selected from the pretrained GFM vocabulary according to token frequency.
The protocol contains three key components:

\begin{itemize}
  \item \textbf{Task-Adaptive Codebook.} The protocol includes a compact set of protocol tokens, each corresponding to a semantic primitive such as an object, action, state, or spatial concept. Compared with the general-purpose GFM tokenizer, the task-adaptive codebook retains only task-relevant semantics, thereby substantially reducing the number of source bits required per transmitted token. For example, if the task-adaptive codebook contains no more than $32$ protocol tokens, each protocol-token ID requires only $5$ source bits, i.e., $\lceil \log_2 32 \rceil = 5$. At the same time, each protocol token is associated with a precise meaning, which suppresses variability and ambiguity in expression. In a collaborative object-transport task, for example, the codebook may include protocol tokens for key objects (e.g., ``apple'' and ``banana''), actions (e.g., ``grasp'' and ``move''), and states (e.g., ``found'' and ``clear''). By restricting the codebook to these essential task-relevant elements, agents can communicate more efficiently and with less ambiguity.
  
  \item \textbf{Syntax Rules.} The protocol also specifies formal rules for combining protocol tokens into valid protocol-token sequences. These rules standardize sequence construction and interpretation across agents, ensuring consistency and reducing ambiguity. They also facilitate subsequent decoding. For example, constraining the order of protocol tokens in a sequence helps the receiver identify erased or unreliable protocol tokens under noisy wireless transmission.

  \item \textbf{Examples.} The protocol further provides example mappings between messages and encoded token sequences. These examples enable the agents, specifically their GFMs, to follow the protocol through in-context prompting without requiring additional retraining.
\end{itemize}

\subsubsection{Protocol-Based Encoding and Decoding}
Once the task-adaptive protocol is established, each agent adopts a protocol-based encoding and decoding pipeline during communication. As illustrated in Part C of Fig.\,\ref{fig:framework}, the communication flow includes five stages: message generation, protocol-based semantic distillation, binary mapping of protocol-token IDs, wireless transmission, and semantic reconstruction from received protocol tokens. This pipeline is tightly integrated into the GFM-centered cognition--action loop, so that decoded messages can be directly used by the GFM for subsequent reasoning and action generation.

\begin{itemize}
  \item \textbf{Transmitter Side.} When an agent decides to transmit a message, it first generates a raw natural-language message that conveys the intended semantics but may contain redundant wording or task-irrelevant details. Instead of directly tokenizing this message with the general-purpose tokenizer of the GFM, the agent employs a GFM-based encoder to distill it into several protocol tokens. Specifically, the raw message and the shared protocol are formatted as the input to the GFM, and the output is constrained to contain only codewords defined in the protocol. In this way, the transmitted tokens remain semantically relevant, compact, and fully compliant with the shared protocol.
  
  \item \textbf{Receiver Side.} After wireless transmission, some protocol tokens may be received correctly, erased, or marked as unreliable due to channel noise. The receiver therefore combines the received tokens with the shared protocol and its local memory in a GFM-based decoder to recover the intended semantics. The decoder converts the received tokens, whether clean or corrupted, into a message suitable for downstream reasoning. For example, in Part C of Fig.\,\ref{fig:framework}, a protocol token corresponding to a room name is corrupted during transmission. By leveraging the protocol together with contextual memory, such as dialogue history and action history, the receiver can infer that the missing token denotes a location and estimate the transmitter's likely current room.
\end{itemize}

Overall, the proposed framework shifts TokCom from general-purpose token exchange to protocol-guided, task-oriented agent communication. The transmitted protocol tokens can therefore be regarded as an interpretable agent language tailored to wireless CEAI.

\begin{figure*}
  \centering
  \includegraphics[width=0.95\textwidth]{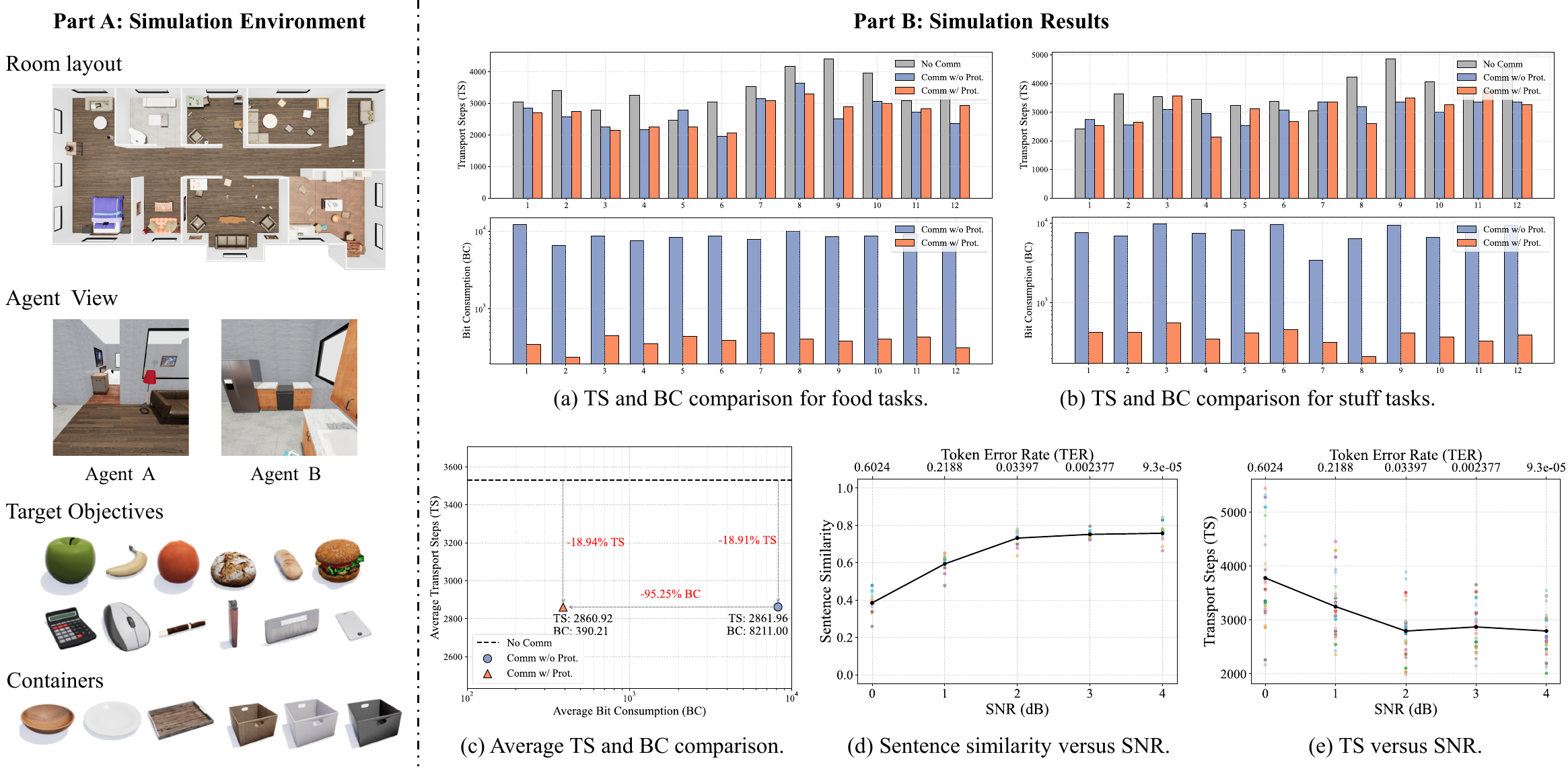}
  \caption{Simulation environment and results. Part A shows the room layout, target objects, and containers. Parts B(a) and B(b) report TS and BC for food and stuff tasks, respectively. Part B(c) compares average TS and BC across schemes. Parts B(d) and B(e) show sentence similarity and TS versus SNR.}
  \label{fig:simulation}
\end{figure*}

\section{Case Study}
\subsection{Scenario Setup}
\textbf{Environment and Task:} We evaluate the proposed framework on the ThreeDWorld object-transport benchmark~\cite{ICRA/22/Gan}. 
Fig.\,\ref{fig:simulation} Part A shows a representative simulation environment, including the multi-room house layout, target objects, and optional containers. Two embodied agents collaboratively search for these objects and transport them to a designated goal location, optionally using the containers to improve efficiency.
Both agents have identical capabilities, including egocentric RGB-D perception, autonomous navigation, object manipulation, and message transmission. 
Following the modular design in~\cite{ICLR/24/Zhang}, RGB-D observations are converted into text by a perception module, and this conversion is assumed to perfectly preserve the semantics of the RGB-D images. The same design can be extended to arbitrary modalities by converting each modality into text through a similar module.
They are initialized at different locations with partially observable fields of view, making communication useful for coordinating exploration and avoiding redundant or conflicting actions. We evaluate $24$ task instances, including $12$ food scenarios and $12$ stuff scenarios.

\textbf{Model and Communication Setup:} We use the instruction-tuned Gemma 4 31B Dense model\footnote{Google DeepMind, ``Gemma 4 Model Card,'' \url{https://ai.google.dev/gemma/docs/core/model_card_4}, accessed Jun. 17, 2026.} as the text-only GFM backbone for agent reasoning and communication. 
The model is frozen and used only for inference. It is neither trained from scratch nor fine-tuned. Protocol initialization, encoding, and decoding are realized by in-context prompting with the default sampling configuration. We deploy the model on two NVIDIA GeForce RTX 3090 GPUs, where the measured generation throughput is $35.83$ output tokens/s and the GPU memory footprint is about $45.4$\,GB. These figures depend on the hardware, quantization scheme, and cache configuration, and they characterize our setup rather than hardware-independent properties of the method. A lighter GFM capable of following the shared protocol can replace the current backbone to reduce computational cost.
At each simulation step, the GFM selects from feasible high-level actions, such as searching the current room, picking up an object, navigating to a location, or sending a message. 
For Conventional TokCom, natural-language messages are directly tokenized by Gemma 4's default tokenizer, and the resulting token indices are mapped into binary payloads for transmission. 
For Protocol-Guided TokCom, agents initialize a task-adaptive protocol at the beginning of each task and exchange compact protocol-token IDs instead of full vocabulary indices. 
Both schemes use fixed-length token-ID mapping without entropy coding. In the noisy-channel evaluation, they share the same digital communication chain, consisting of a rate-1/2 convolutional code, quadrature phase shift keying (QPSK) modulation, and soft-decision Viterbi decoding.

\textbf{Evaluation Metrics:} All evaluated task instances are completed within the maximum step budget. We therefore focus on task efficiency and communication efficiency using two metrics:
\begin{itemize}
    \item \textit{Transport Steps (TS):} The cumulative number of simulation steps required to transport all target objects to the designated goal, where fewer steps indicate higher task efficiency.
    \item \textit{Bit Consumption (BC):} The total number of source-side payload bits generated by tokenizing or protocol-encoding all agent-to-agent messages before channel coding, excluding channel-coding redundancy and packet headers. Protocol initialization overhead is not included in BC, since the generated protocol can be reused over many communication rounds. BC therefore measures subsequent message payloads after a shared protocol is available. When agents use the same GFM with a fixed random seed and a common sampling configuration, they can generate an identical protocol locally from the shared task context, so the protocol itself need not be transmitted. This excluded overhead becomes more significant for short tasks or frequent protocol regeneration.
\end{itemize}

\textbf{Baselines and Proposed Method:} We compare three schemes:

\begin{itemize}
  \item \textbf{No Comm.:} Agents operate independently without any inter-agent communication, serving as a non-communication reference.

  \item \textbf{Conventional TokCom:} Agents generate natural-language messages with the GFM and directly transmit the corresponding pretrained-token indices. Upon reception, the recovered message is fed into the receiver's GFM for inference.
  
  \item \textbf{Protocol-Guided TokCom (Proposed):} Agents initialize a task-adaptive protocol, distill natural-language messages into compact protocol-token sequences, and reconstruct actionable semantics from received protocol tokens.
\end{itemize}

Both Conventional TokCom and Protocol-Guided TokCom are evaluated without entropy coding. Entropy coding reassigns code lengths to existing tokens rather than changing the semantic representation, and it can be applied to either token stream.

\subsection{Performance Analysis}
Under error-free communication, Fig.\,\ref{fig:simulation} Parts B(a) and B(b) compare TS and BC for the food and stuff tasks, respectively. Communication generally reduces TS by helping agents coordinate exploration and avoid redundant actions, such as searching the same room or attempting to grasp the same object. In a few scenarios, communication-enabled agents require slightly more steps than the No Comm. reference, mainly due to the stochasticity of GFM-based decisions under partial observability. On average, Conventional TokCom reduces TS by $21.02\%$ for food tasks and $16.44\%$ for stuff tasks compared with No Comm.

Protocol-Guided TokCom achieves step reductions comparable to Conventional TokCom, suggesting that the task-adaptive protocol preserves coordination-critical semantics. In preliminary trials, simply asking a GFM to compress messages without a shared protocol did not reliably improve TS. Although the transmitted text became shorter, compressed expressions generated by one agent could be ambiguous to another agent, because the agents did not share an explicit codebook or syntax for interpreting the compressed tokens. In contrast, Protocol-Guided TokCom constrains both agents to a common set of semantic primitives and composition rules, making compressed messages more consistently interpretable.

The communication saving comes from semantic distillation and compact protocol indexing. Semantic distillation reduces the number of transmitted tokens by removing task-irrelevant wording, while compact protocol indexing maps each protocol token to a small ID instead of an index in the full GFM vocabulary. Without the protocol, agents transmit an average of $476.08$ tokens per episode; with the protocol, this drops to $78.88$ tokens, corresponding to a reduction of $83.43\%$. Since the task-adaptive codebook contains only about $22$ to $28$ protocol tokens, each protocol-token ID requires $5$ bits, compared with $18$ bits for Gemma 4's general-purpose tokenizer. Together, these two factors reduce the total source payload to about $5\%$ of that required by Conventional TokCom while achieving comparable transport efficiency, as shown in Fig.\,\ref{fig:simulation} Part B(c).

We further conduct a preliminary noisy-channel evaluation over an additive white Gaussian noise (AWGN) channel. Following the token-erasure abstraction used in TokCom studies~\cite{MWC/25/Qiao}, protocol-token IDs detected as unreliable after channel decoding are replaced by \texttt{[MASKED]}, so this evaluation characterizes robustness to token erasures. In practical systems, such erasures can be obtained through token-level cyclic redundancy checks, confidence-based demodulation, or packet-level error detection. 
Undetected substitutions would be more challenging than erasures, because the received tokens are not marked as unreliable. Guided by the shared protocol and context, the GFM may still detect or correct some inaccurate tokens, but task performance may degrade. Both schemes adopt the same masking strategy. The token error rate (TER) is computed as the fraction of transmitted protocol-token IDs marked unreliable before GFM-based semantic reconstruction.

The protocol structure makes erasure recovery more constrained. Since valid tokens come from a compact task-adaptive codebook, the decoder searches over a much smaller protocol-token space than the full GFM vocabulary. Syntax rules and neighboring protocol tokens can further indicate the missing token type, such as object, action, location, or state. The receiver can then combine this type constraint with local memory, dialogue history, and action history to infer the missing semantics more effectively.

Fig.\,\ref{fig:simulation} Parts B(d) and B(e) show sentence similarity and TS versus signal-to-noise ratio (SNR). As the SNR increases from $0$\,dB to $2$\,dB, TER drops sharply from $0.60$ to $0.03$, and TS decreases accordingly. When the SNR reaches $2$\,dB or higher, the average TS stabilizes and matches the error-free case. This suggests that exact sentence recovery is not always necessary, as long as coordination-critical semantics can still be inferred. Overall, the case study illustrates the potential robustness of task-adaptive protocols for wireless CEAI.

\section{Future Research Directions}
\subsection{Joint Source-Channel Coding}
Due to the discrete nature of codebooks, TokCom relies on discrete protocol-token ID transmission, which poses challenges in low-SNR regimes. Joint source-channel coding (JSCC) improves robustness through end-to-end training by jointly optimizing source and channel encoding. Future work should explore lightweight neural networks adapted to specific communication scenarios, incorporating task information, protocol constraints, and channel statistics to improve resilience in noisy wireless environments. For example, meta-learning can be used to initialize a lightweight JSCC model that can rapidly adapt to new tasks and channel conditions with minimal additional training.

\subsection{Dynamic Protocol Adaptation}
Agents may discover new objects during task execution that are unanticipated during protocol initialization. Thus, a dynamic protocol update mechanism is essential for sustained collaboration. 
Future research should investigate methods for efficiently integrating newly discovered semantic primitives into the current protocol with minimal overhead, while maintaining coherent multi-agent coordination across environmental changes. 

\subsection{Heterogeneous Agent Interoperability}
The proposed framework assumes that agents use compatible GFM implementations for encoding and decoding. However, heterogeneous agents may employ different model architectures, training procedures, or quantization schemes. Future work should investigate protocol mechanisms enabling interoperability across diverse GFM implementations, develop robust decoding strategies tolerant to model variations, and establish formal verification methods that ensure semantic consistency across heterogeneous agent populations.

\section{Conclusion}
This article investigated TokCom-assisted CEAI for efficient and task-effective collaboration among embodied agents over wireless networks. We first discussed how TokCom supports CEAI through insight sharing, intent alignment, and interactive control. To address the overhead and ambiguity of directly transmitting pretrained GFM tokens, we proposed a task-adaptive protocol-guided framework with compact protocol tokens, syntax rules, and GFM-based semantic distillation and reconstruction. The case study suggests that protocol-guided TokCom can substantially reduce source payload bit consumption while largely preserving collaborative task performance under token erasures. Future advances in channel-aware token coding, dynamic protocol adaptation, and heterogeneous GFM interoperability will be important for practical deployment.

\bibliographystyle{IEEEtran}
\bibliography{References}

\end{document}